\documentclass[prl,showpacs,superscriptaddress,twocolumn,nobibnotes]{revtex4-2} 

\usepackage{amsmath}
\usepackage{amssymb}
\usepackage{amsthm}
\usepackage{bm}
\usepackage{comment}
\usepackage{float}
\usepackage{hyperref}
\usepackage{xr-hyper}
\usepackage{titlesec}
\usepackage[T1]{fontenc}
\usepackage{mathtools} 
\usepackage[dvipsnames]{xcolor}
\usepackage[autostyle, english = american]{csquotes}
\usepackage{graphicx}

\usepackage[ruled,vlined,linesnumbered]{algorithm2e}

\hypersetup{
  colorlinks   = true,
  urlcolor     = blue,
  linkcolor    = black,
  citecolor   = black 
}

\begin{document}

\title{Vortex rings shape the feeding-efficiency landscape of jellyfish} 

\author{Harry Walden}
\thanks{These authors contributed equally to this work.}
\affiliation{Department of Mathematics, Massachusetts Institute of Technology, Cambridge, MA 02139, USA}

\author{Ashutosh Tripathi}
\thanks{These authors contributed equally to this work.}
\affiliation{Department of Mathematics, Massachusetts Institute of Technology, Cambridge, MA 02139, USA}

\author{Teo Lara}
\thanks{These authors contributed equally to this work.}
\affiliation{Department of Mathematics, Massachusetts Institute of Technology, Cambridge, MA 02139, USA}

\author{Brandon Weissbourd}
\affiliation{Department of Biology, Massachusetts Institute of Technology, Cambridge, MA 02139, USA}
\affiliation{Picower Institute for Learning and Memory, Massachusetts Institute of Technology, Cambridge, MA 02139, USA}

\author{Jörn Dunkel}
\affiliation{Department of Mathematics, Massachusetts Institute of Technology, Cambridge, MA 02139, USA}

\begin{abstract}
Fluid flows enable and limit the feeding and locomotion of sea creatures across a wide range of scales. Although much is known about how microorganisms navigate life at low Reynolds numbers, the effect of inertial flows on the feeding-efficiency of larger marine organisms remains less well understood. Here, using jellyfish as an archetypal example, we study the trade-off between prey capture rate and energy expended to generate inertial flows by surface deformations. We develop a feeding-efficiency landscape that predicts an optimal feeding regime at the transition from Stokes pumping to inertial prey transport, mediated by a self-generated vortex ring. These predictions provide a unifying biophysical explanation of size and beat frequencies reported for jellyfish species from different taxa, with vortex-ring-dominated transport supported by our experimental observations for the moon jellyfish \textit{Aurelia aurita}. More broadly, our results demonstrate how centimetric marine predators can exploit inertial flows for efficient prey capture.
\end{abstract}

\maketitle
\begingroup
\renewcommand\thefootnote{}
\endgroup

Marine life depends on fluid flows for locomotion, prey capture, and nutrient transport~\cite{Brennen1977, Omori2025, Costello1994, peng_transport_2009, Kirkegaard2016, Goldstein2015, Liu2025}. Aquatic microbes~\cite{Omori2025, Liu2025}, sponges~\cite{Asadzadeh2020} and corals~\cite{Shapiro2014} typically experience viscosity-dominated reversible Stokes flow~\cite{purcell_life_1977}, whereas larger, faster-moving sea animals create more complex inertial flows~\cite{dawoodian_kinetics_2021, 2005_jellyfish_experimental_dabiri, peng_transport_2009}. Operating in a transition regime where both viscous and inertial effects are relevant~\cite{Cartwright_jellyfish_fossil,pallasdies_single_2019}, 
jellyfish are among the first multicellular life forms to use their musculature to generate flows for locomotion and predation~\cite{Cartwright_jellyfish_fossil,pallasdies_single_2019}. Beyond their ecological importance to the marine ecosystem~\cite{Hays2018trophic,Purcell_jelly_blooms_and_humans_2012,Brotz_2012_jelly_humans_population}, jellyfish have emerged as a model system for investigating the interplay between active body deformations and fluid dynamics, attracting increasing interest in bio-inspired soft robotics~\cite{Nawroth2012A, Anuszczyk2026,Anuszczyk2024Hybrid} and neuromechanical actuation~\cite{weissbourd_genetically_2021}. Recent experimental~\cite{2005_jellyfish_experimental_dabiri, McHenry_Aurita_data_2003, 2002_swimming_Colin, gemmell_control_2015, Anuszczyk2026} and computational~\cite{hoover_numerical_2015, hoover_quantifying_2017, Hamlet_jelly_bell_numerical_2011, Hoover_Porras_Miller_resonance_swimming_2019} studies have revealed important insights into the fluid mechanics of jellyfish swimming. By contrast, less is known about how jellyfish can utilize self-generated flows for predation and feeding. Here, aiming to understand more broadly how marine life can sustain itself at intermediate Reynolds numbers, we consider jellyfish as an archetypal model organism to investigate the physical limits of prey capture and, hence, energy consumption at the transition to inertial flow.
\par
\begin{figure*}[t]
    \includegraphics{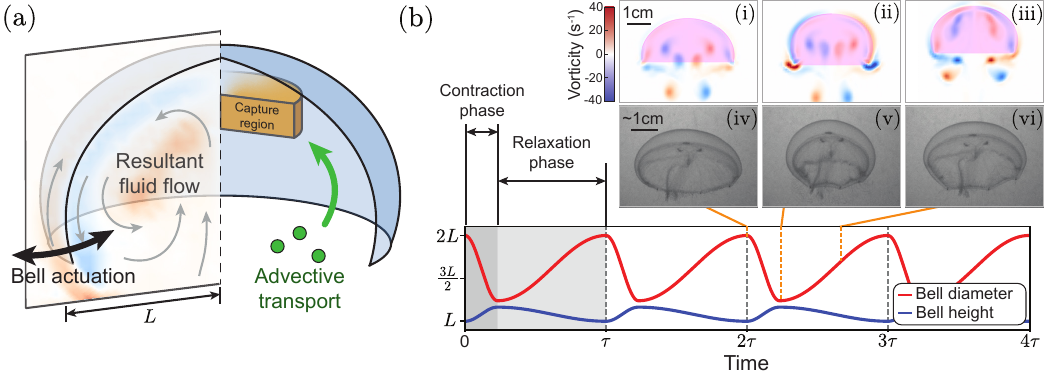}
    \caption{Jellyfish feeding couples periodic bell motion to prey transport. (a) Schematic of the jellyfish feeding process: Bell actuation generates a fluid flow that transports prey towards the capture region. (b) Bell deformations are characterized by a fast contraction phase followed by slow relaxation, shown here in the model~(i--iii), with in-plane vorticity~($\text{Re}=400$, see Supplementary Video S2), and experimentally~(iv-vi). The evolution of the bell radius and height is designed to match experimental observations of~\emph{A.~aurita} $(\tau\sim 1\,\mathrm{s})$~\cite{McHenry_Aurita_data_2003}.}
    \label{fig: schematic}
\end{figure*}

Jellyfish induce flow through muscular contractions of their bell-shaped bodies, generating flows that carry prey towards the capture region at the mouth~\cite{Wagner2020,dawoodian_kinetics_2021}~(Fig.~\ref{fig: schematic}a,b). To quantify prey capture rates and estimate the energetic cost of generating these flows, we study a continuum model that accounts for fluid-structure interactions and advective prey transport. Combining fluid simulations with  metabolic conversion estimates, we construct a feeding-efficiency landscape that predicts the power generation from a jellyfish's body size and its characteristic Reynolds number. The optimal feeding regime predicted by our analysis coincides with the  transition  from viscous to inertial flow, and is consistent with previously reported body size and feeding-stroke parameters for various jellyfish species~\cite{Monfort_jelly_data_2023, McHenry_Aurita_data_2003, 2002_swimming_Colin}. In particular, the model implies that centimetric jellyfish can achieve optimal feeding-efficiency by generating vortex rings, in agreement with our experimental observations for the moon jellyfish \textit{Aurelia aurita}.

We develop a hydrodynamic transport model by representing the jellyfish as a hemiellipsoidal bell undergoing axisymmetric deformations in a viscous, incompressible fluid. A typical deformation sequence matched~[SI] to \emph{A.~aurita}~\cite{McHenry_Aurita_data_2003, dawoodian_kinetics_2021} is shown in Fig.~\ref{fig: schematic}b. The fluid velocity field, $\mathbf{u}(\mathbf x, t)$, obeys the Navier--Stokes equations in the bulk and satisfies far-field decay conditions and no-slip boundary conditions on the surface of the jellyfish. The net motion of the jellyfish is self-consistently determined by force and torque balance~\cite{Childress_1981_SwimmingFlying,hoover_numerical_2015}. Assuming that prey transport is dominated by advection,
 the effective prey concentration, $c(\mathbf{x},t)$, evolves according to
 \begin{equation}
        \frac{\partial c(\mathbf{x},t)}{\partial t}+\mathbf{u}(\mathbf{x},t)\cdot{\nabla}c(\mathbf{x},t)=0
\label{eq: concentration advection eq}
\end{equation}
with an absorbing boundary condition  $c(\mathbf{x},t)=0$ in the  prey capture (\lq mouth\rq) region $M(t)$, which is co-moving with the body frame and acts as a sink for the auxiliary prey field~[SI]~(Fig.~\ref{fig: schematic}a). Typical prey, such as fish eggs, larvae, and small mesozooplankton, can be sparsely distributed~\cite{Han_Aurita_weight, Stoecker_aurita_feeding_zooplankton, Zooplanktonenergydensity}; in this case, Eq.~\eqref{eq: concentration advection eq} provides a statistical description for long-run averages of prey capture rates by modeling the initial prey distribution as a Poisson point process with intensity $c(\mathbf{x},0)$~[SI].

Focusing on jellyfish that exhibit shape dynamics similar to \emph{A.~aurita} (Fig.~\ref{fig: schematic}b), the prey transport is essentially determined by only two parameters, the bell radius $L$ and the contraction period $\tau$ (e.g., $L\sim 1.5\,\text{cm}$ and $\tau \sim 1\,\text{s}$ for \emph{A.~aurita}~\cite{McHenry_Aurita_data_2003}). These can be combined with the density $\rho$ and dynamic viscosity $\mu$ of water to form the characteristic Reynolds number,~$\text{Re} \equiv \rho L^2/\mu \tau$, which measures the relative importance of inertial and viscous forces during a feeding stroke. By comparing viscous dissipation losses at different values of $\text{Re}$ with energy gains due to metabolic energy conversion of prey, we can next identify optimal feeding regimes.

\begin{figure*}
    \includegraphics{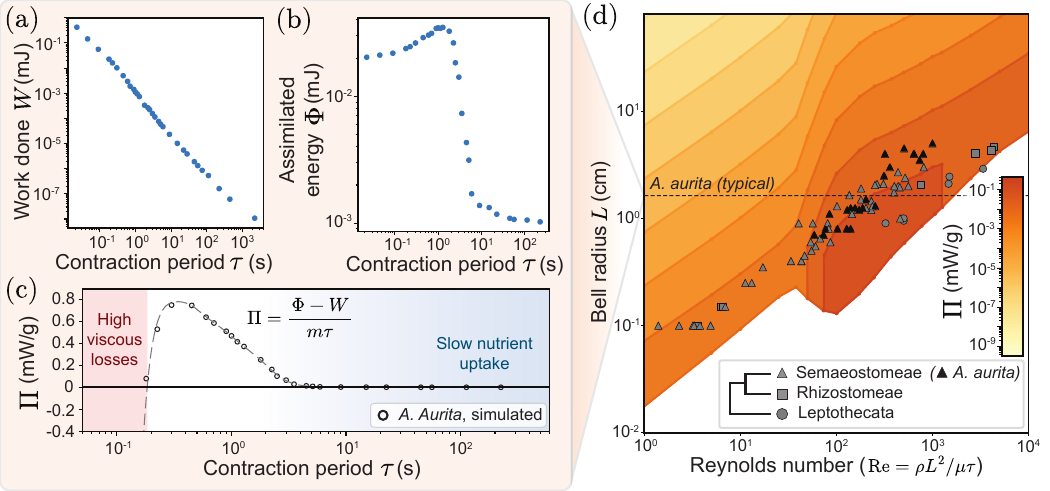}
    \caption{Energy balance analysis predicts optimal feeding regimes relative to hydrodynamic work in agreement with experimental data for various bell-shaped jellyfish species. (a)~Simulations indicate net hydrodynamic work $W$ per contraction decreases with increasing stroke period $\tau$. (b) Assimilated nutrient energy $\Phi$ from feeding exhibits a step-like transition as a function of contraction period. (c) Specific power $\Pi$ is maximized at intermediate stroke periods as viscous losses dominate at short periods, whereas prey capture is limited by stroke rate at long periods (dashed line fitted as a visual aid). Panels (a--c) use  $L=1.5\,\text{cm}$ typical of \emph{A. aurita}. (d) Experimental observations of jellyfish species with shapes similar to \emph{A.~aurita} lie close to the optimal energy balance regime. Positive net specific power gain $\Pi>0$ distinguishes viable (orange) from  unviable (white) regimes. Data aggregated from observational studies~\cite{Monfort_jelly_data_2023, McHenry_Aurita_data_2003, 2002_swimming_Colin} are shown as symbols, labeled by phylogenetic order.}
    \label{fig: energetics}
\end{figure*}

To compute the energetic costs and metabolic gains per feeding stroke, we solve the coupled fluid-prey system via an immersed boundary method~\cite{Peskin_IB_2002} using the adaptive-mesh solver IBAMR~\cite{Griffith_2007_IBAMR,Griffith_2017_IBAMR_validification}~[SI] (Fig.~\ref{fig: schematic}b, i--iii; Video~S2). By integrating the rate of viscous dissipation plus the kinetic energy change in the fluid, we find the total hydrodynamic work $W$ done by the jellyfish over a stroke of duration $\tau$,
\begin{equation}
    W = \int_{0}^{\tau} dt\,\int_{\Omega} d^3 \mathbf{x}\, \left( \mu\|\nabla \mathbf{u}\|^2+ \frac{1}{2}\rho\frac{\partial}{\partial t}\|\mathbf{u}\|^2 \right).
\end{equation}
As illustrated for \emph{A.~aurita} in Fig.~\ref{fig: energetics}a,  $W$ decreases monotonically with $\tau$  consistent with the intuition that slower strokes reduce viscous losses. 

Jellyfish need to capture and metabolize prey to sustain themselves and perform work. The nutrient energy assimilated per stroke cycle, $\Phi$, is proportional to the integrated flux of prey entering the capture region,
\begin{equation}
    \Phi= \gamma \int_{0}^{\tau}dt\, \int_{\partial M(t)}d \mathbf{S}\cdot \big[\mathbf{u}(\mathbf{x},t)-{\mathbf{U}}(t)\big] c(\mathbf{x},t),
    \label{eq: phi integral}
\end{equation}
where $\mathbf U(t)$ is the net velocity of the jellyfish. The prey field $c(\mathbf x,t)$  is obtained by solving Eq.~\eqref{eq: concentration advection eq} with constant initial density $c_0$~$\sim3\times10^{-5}\text{mg}\,\text{cm$^{-3}$}$~[SI]. The metabolic conversion factor $\gamma$ is given by the product of the assimilation efficiency~($\sim0.35$) and prey energy density~(\hbox{$\sim 10^{4}\text{mJ}\,\text{mg$^{-1}$}$)}; see [SI] for detailed metabolic data~\cite{assimilation_efficiency_2,assimilation_efficiency_3,Periera_2014,Han_Aurita_weight,Stoecker_aurita_feeding_zooplankton,Zooplanktonenergydensity,mesozooplanktondensitynewzealnd,mesozooplanktondensity_cyprus,AAuritaRespirationRate3} and sensitivity analyses. Crucially, our simulations show that the assimilated nutrient energy per stroke, $\Phi$, exhibits a step-like transition from a high-energy plateau at short contraction periods to a low-energy plateau at long contraction periods~(Fig.~\ref{fig: energetics}b).

Long-term sustainability of a feeding strategy relies on energy gains $\Phi$ outweighing the cost $W$ of executing a stroke, with the excess devoted to baseline metabolic costs to maintain tissue, for growth and other essential life processes. Allometric scalings predict basal metabolic rate from body weight across the animal kingdom~\cite{West1997Allometric}, including jellyfish, where metabolic demand scales with the body weight of the organism~\cite{Allometric_1987,AAuritaRespirationRate4}. This motivates us to study the net specific power generation,
\begin{equation}
  \Pi\equiv\frac{\Phi - W}{m\tau},
  \label{eq: specific_power}
\end{equation}
where $m$ is the dry mass of the jellyfish, which can be estimated via an empirical scaling with body size~\cite{AAuritaRespirationRate3}~[SI]. We thus contrast absolute power output to the basal metabolic rate, allowing us to compare individuals across disparate length scales.

For \emph{A.~aurita} ($L\sim 1.5$ cm), the specific power $\Pi$ reaches a maximum at an intermediate stroke period comparable to the organism's typical value $\tau~\sim 1$\,s~(Fig.~\ref{fig: energetics}c). In the limit of rapid strokes ($\tau \to 0$), the work $W$ required to execute the stroke eventually exceeds the assimilated nutrient energy~$\Phi$ (Fig.~\ref{fig: energetics}c, left), making fast swimming unviable ($\Pi<0$) in the long term. Conversely, in the limit of slow strokes ($\tau \to \infty$), the mechanical work $W$ is  negligible compared to the assimilated nutrient energy $\Phi$ (Fig.~\ref{fig: energetics}c, right), but the decreased stroke rate also dramatically reduces the prey capture rate. Such a slow-motion strategy will eventually fail to capture enough prey to sustain other life processes. The results in Fig.~\ref{fig: energetics}a-c thus suggest that \emph{A.~aurita} feeds with a stroke which balances prey capture against hydrodynamic work.

\begin{figure*}[t]
    \centering
    \includegraphics[width=1.0\linewidth]{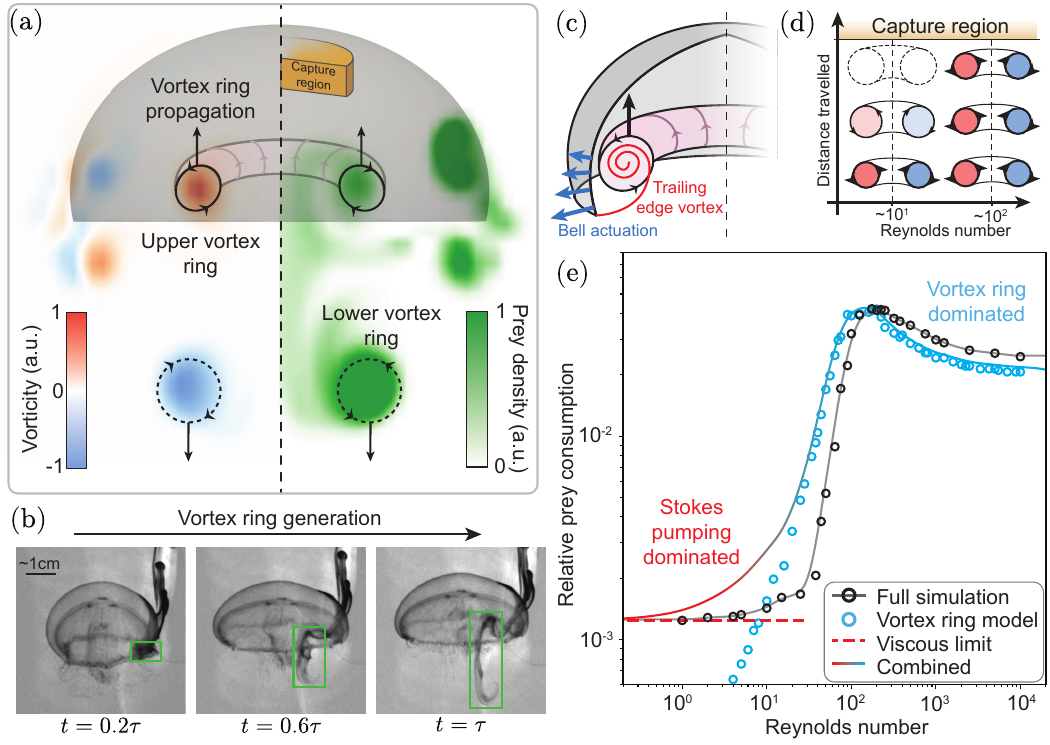}
    \caption{Vortex ring dynamics recover the step-like prey capture transition at high Reynolds number. (a) Fluid-prey modeling predicts the presence of long-lived vortical structures. A snapshot at $\text{Re}=10^3$ shows simulated vorticity (red/blue, left) and prey density (green, right), overlaid with sketches of vortex rings. See Supplementary Video S1 and Supplementary Figure S5 for detailed vortex ring formation. (b) This prediction is borne out experimentally as \emph{A. aurita} interacts with a passive dye. Green boxes indicate dye transport over the course of a contraction, qualitatively matching the numerical predictions [panel (a)]. See Supplementary Video S3. (c) Schematic showing the formation of the upper vortex ring during bell expansion and the subsequent propagation. (d) At high Reynolds numbers, vortex rings become long-lived and play a key role in prey transport. The schematic illustrates the decay of vortical structure before reaching the capture region at $\text{Re}=10^1$, while at $\text{Re}=10^2$ the vortex ring remains coherent. (e) The minimal vortex-ring model recovers the phenomenology of the full simulation [Fig.~\ref{fig: energetics}]. The vortex ring model and fitted Stokes contribution are combined to reproduce the step-like feeding transition observed in solutions to the full immersed boundary problem.}
    \label{fig: vortex ring dynamics}
\end{figure*}

To test whether the above results generalize to  other jellyfish species, we considered bell radii $L$ ranging from sub-millimeters to meters. The heatmap in Fig.~\ref{fig: energetics}d distinguishes unviable (white, $\Pi<0$) from viable (orange, $\Pi>0$) parameter regimes in the ($\text{Re},L$) parameter plane. By overlaying typical ($\text{Re},L$)-values (symbols) derived from previous experimental studies~\cite{McHenry_Aurita_data_2003, Monfort_jelly_data_2023,2002_swimming_Colin} for bell-shaped jellyfish covering a wide range of sizes and taxa, one observes that most of them lie close to the optimal feeding regime (dark-orange) predicted by our model. Sensitivity analysis~[SI] shows that reduced prey availability shifts the optimal regime to lower Reynolds number. Yet, even as the predicted optimum shifts under such more adverse conditions, the experimental data remain in the viable region. This may reflect that jellyfish adopt a feeding strategy that is both energy efficient and robust to changes in nutrient availability.

To understand the dominant fluid-dynamical mechanisms in the optimal jellyfish feeding regime (Fig.~\ref{fig: energetics}d), we study the vorticity field  $\nabla\times \mathbf{u}$. At (Re,$L$) values typical of \emph{A. aurita}, our simulations predict the formation of a vortex ring pair~(Fig.~\ref{fig: vortex ring dynamics}a; Video S2). We confirm this prediction qualitatively by imaging the transport of a passive tracer by \emph{A. aurita}~(Fig.~\ref{fig: vortex ring dynamics}b)~[SI, Video S3]. 
Previous studies of vortex rings in jellyfish swimming~\cite{2005_jellyfish_experimental_dabiri,Gemmell2014vortex,gemmell_control_2015} showed that the back-reaction against the \emph{lower} vortex ring, formed by the initial bell contraction during a stroke, supports efficient propulsion. By contrast, we find that advective prey transport to the mouth prey is dominated by the \emph{upper} vortex ring, which is formed during the relaxation phase of a stroke~(Fig.~\ref{fig: vortex ring dynamics}c). Our results are consistent with \emph{in silico} and \emph{in vivo} studies that used Lagrangian coherent structures to identify domains of prey capture for jellyfish~\cite{peng_transport_2009, dawoodian_kinetics_2021}, while also unifying the fluid dynamics with biological data. 

We also obtain higher resolution over Reynolds number, which allows us to test the hypothesis that vortex rings enable efficient feeding (Fig.~\ref{fig: vortex ring dynamics}d). Our simulations predict a steep transition from inefficient prey capture at low Re values to high prey capture rates at high Re values (black circles in Fig.~\ref{fig: vortex ring dynamics}e). 
To confirm that the upper vortex ring indeed accounts quantitatively for the observed transition, we formulate a reduced model consisting of a single vortex-ring propagating through a cylinder of fluid bounded by a rigid wall. Following~\cite{Didden_vortex_ring_circulation}, the initial circulation $\Gamma$ is set by the scaling relation $\rho\Gamma/\mu \sim \text{Re}$. By computing the resulting fluid flow via a fixed-grid numerical method~[SI], we find that the prey consumption in the reduced vortex-ring model~(Fig.~\ref{fig: vortex ring dynamics}e, blue circles) agrees well with the full fluid-prey simulations~(Fig.~\ref{fig: vortex ring dynamics}e, black circles) both at and above the transition~($\text{Re}>10^1$). Below the transition, the reduced model becomes inaccurate, predicting a decaying prey capture rate in the low Re~(Stokes) limit, in contrast to the plateau observed in the full model simulations. This discrepancy is explained by the fact that, although advection cannot generate net transport over a full cycle~\cite{purcell_life_1977} as $\text{Re}\to 0$, the absorbing capture region breaks time-reversibility, leaving a finite capture fraction. We thus refer to this mode of prey capture as Stokes pumping. By approximating the total prey consumption as the sum of the asymptotic Stokes pumping contribution (red-dashed line) and vortex-mediated contributions (blue circles), one recovers qualitatively (solid red-to-blue curve, Fig.~\ref{fig: vortex ring dynamics}e) the transition behavior seen in the full model, confirming that the upper vortex ring dominates prey capture at sufficiently high Reynolds numbers.

Jellyfish have attracted an increasing interest in recent decades in the context of biomaterials~\cite{alexander_visco-elastic_1964,Megill2005elasticity,hoover_quantifying_2017}, soft robotics~\cite{Nawroth2012A, Anuszczyk2026,Anuszczyk2024Hybrid,Villanueva_2011,villanueva_flexible_2014,Gunnarson2025Surfing,Wang2023Robot}, and  neuromechanics~\cite{weissbourd_genetically_2021}. Here, we combined numerical fluid-transport analysis with metabolic estimates and experimental observations to show how these organisms can utilize self-generated inertial flows for efficient feeding. The fact that the optimal feeding regime predicted by our fluid-prey model agrees with experimental observations for jellyfish species spanning a wide range of sizes and belonging to different taxa highlights the importance of physical selection pressures during the evolution of biological systems.

\paragraph{Acknowledgments}
\textit{Aurelia aurita} were acquired from the New England Aquarium. T.L., H.W., and A.T. acknowledge the support of the MIT Mathematics SPUR program. This research received support through the MathWorks Science Fellowship (H.W.). B.W. was supported by the Picower Institute for Learning and Memory, The Freedom Together Foundation, and The Searle Scholars Program. This research received support through Schmidt Sciences, LLC (to J.D.), and the MathWorks Professorship Fund (to J.D.). 
 
\nocite{dedalus_citation_burns}
\nocite{Mahajan_zernike_polynomials}
\nocite{Periera_2014}
\nocite{Han_Aurita_weight}
\nocite{Stoecker_aurita_feeding_zooplankton}
\nocite{Zooplanktonenergydensity}
\nocite{mesozooplanktondensitynewzealnd}
\nocite{mesozooplanktondensity_cyprus}
\nocite{assimilation_efficiency_2}
\nocite{assimilation_efficiency_3}
\nocite{AAuritaRespirationRate3}

\bibliography{refs}

\end{document}